\documentclass[reprint,nofootinbib,floatfix,aps,prd]{revtex4-2}

\AtBeginDocument{\RenewCommandCopy\qty\SI}

\usepackage{graphicx}
\usepackage{amsmath,amssymb,mathtools}
\usepackage{bm}
\usepackage{siunitx}
\usepackage{xcolor}
\usepackage[table]{xcolor}

\usepackage[caption=false]{subfig}

\usepackage{booktabs}
\usepackage{dcolumn}
\usepackage{multirow}
\usepackage{array}
\usepackage{adjustbox}
\usepackage{rotating}

\usepackage{orcidlink}
\usepackage[T1]{fontenc}

\usepackage{hyperref}

\newcommand{\lcdm}{\texorpdfstring{$\Lambda$CDM}{LCDM}}
\newcommand{\olcdm}{\texorpdfstring{$o\Lambda$CDM}{oLCDM}}
\newcommand{\nulcdm}{\texorpdfstring{$\nu\Lambda$CDM}{nuLCDM}}
\newcommand{\wowacdm}{\texorpdfstring{$w_0w_a$CDM}{w0waCDM}}

\newcommand{\nuwowacdm}{\texorpdfstring{$\nu w_0 w_a$CDM}{nuw0waCDM}}
\newcommand{\Planck}{\textit{Planck}}

\newcommand{\Om}{\Omega_\mathrm{m}}

\newcommand{\Ok}{\Omega_k}
\newcommand{\OL}{\Omega_\Lambda}
\newcommand{\ode}{\Omega_{\mathrm{de}}}
\newcommand{\Obhtwo}{\Omega_\mathrm{b}h^2}
\newcommand{\Ocdmhtwo}{\Omega_\mathrm{cdm}h^2}

\newcommand{\Neff}{N_{\mathrm{eff}}}

\newcommand{\DM}{D_\mathrm{M}}

\newcommand{\sumnu}{\sum m_\nu}

\newcommand{\la}{\ell_a}

\newcommand{\eV}{\,\mathrm{eV}}
\newcommand{\kmsMpc}{\,\mathrm{km\,s^{-1}Mpc^{-1}}}

\newcommand{\cmbthree}{\texttt{CMB-3}}
\newcommand{\cmbthreew}{\texttt{CMB-3$w$}}
\newcommand{\cmbfournu}{\texttt{CMB-4$\nu$}}
\newcommand{\cmbfourk}{\texttt{CMB-4k}}
\newcommand{\cmbfournuw}{\texttt{CMB-4$\nu w$}}

\begin{document}

\preprint{000-000-000}

\title{CMBComp: A Simple and Accurate Compressed CMB Likelihood for\texorpdfstring{\\}{ }
       Dark Energy, Curvature, and Massive Neutrinos}

\author{Amogh Srivastav\,\orcidlink{0009-0006-1163-1600}}
\email{23b1826@iitb.ac.in}
\affiliation{Department of Physics,
  Indian Institute of Technology Bombay, Mumbai 400076, India}

\author{Prakhar Bansal\,\orcidlink{0009-0000-7309-4341}}
\email{prakharb@umich.edu}
\affiliation{Department of Physics and Leinweber Institute for Theoretical Physics,
  University of Michigan, 450 Church St, Ann Arbor, MI 48109, USA}

\author{Dragan Huterer\,\orcidlink{0000-0001-6558-0112}}
\email{huterer@umich.edu}
\affiliation{Department of Physics and Leinweber Institute for Theoretical Physics,
  University of Michigan, 450 Church St, Ann Arbor, MI 48109, USA}

\date{\today}

\begin{abstract}
We present \textsc{CMBComp}, a compact and accurate compressed cosmic microwave background (CMB) likelihood that captures the dominant geometric information of the full CMB likelihood derived from the combined SPT-3G D1\,+\,ACT~DR6\,+\,\Planck\ PR3 primary CMB anisotropy\,+\,\Planck\ PR4/NPIPE CMB-lensing dataset, which we collectively refer to as SPA. The compression is fast to evaluate and trivial to implement in standard inference pipelines. We construct and validate it in five model spaces: the spatially flat cosmological-constant model (\lcdm), and its dynamical-dark-energy (\wowacdm), massive-neutrino (\nulcdm), non-flat (\olcdm), and joint massive-neutrino--dynamical-dark-energy (\nuwowacdm) extensions. Five compressed likelihoods are introduced, 
%\cmbthree, \cmbthreew, \cmbfournu, \cmbfournuw, and \cmbfourk,
corresponding to two three-parameter compressions for the dark-energy sector (\cmbthree\ for \lcdm, \cmbthreew\ for \wowacdm), two four-parameter compressions for the neutrino sector (\cmbfournu\ for \nulcdm, \cmbfournuw\ for \nuwowacdm), and a four-parameter curvature compression (\cmbfourk), respectively. Combining each compressed likelihood with the DESI DR2 baryon acoustic oscillation (BAO) data, we demonstrate that the resulting posteriors agree to high precision with those obtained from the corresponding full-CMB chains. \textsc{CMBComp} is therefore particularly well suited to cosmological inference for models that modify the late-time expansion history, enabling accurate CMB constraints to be incorporated into new analyses with minimal computational overhead and without reliance on a full Boltzmann-solver-based inference pipeline. The compressed likelihood files and example notebooks accompanying \textsc{CMBComp} are made publicly available at \url{https://github.com/Amoghsriv/CMBComp}.
\end{abstract}

\maketitle

\section{Introduction}
\label{sec:intro}

The cosmic microwave background (CMB) is one of the most informative probes of cosmology, but its full likelihood is computationally expensive and structurally complex to deploy in exploratory analyses. This is particularly limiting for late-time phenomenological cosmologies, for fast scans across large model spaces, and for joint analyses with low-redshift geometric probes such as baryon acoustic oscillations (BAO). In all these settings it is useful to replace the full CMB likelihood with a low-dimensional compression that retains the information most relevant to the background expansion.

The conceptual basis for such a compression has been well established. For smooth dark-energy and geometry-driven extensions of the standard cosmological model, the dominant CMB information for late-time inference is encoded in a small number of geometric quantities: the angular size of the sound horizon at recombination, the comoving distance to the last-scattering surface, and one or two physical density parameters~\cite{Bond:1997wr,Wang:2007mza,Mukherjee:2008kd,Elgaroy:2007bv,Vonlanthen:2010cd,Huang:2015vpa,Zhai:2019nad}. These are conventionally captured by the \emph{shift parameter}~$R$, the \emph{acoustic angular scale}~$\la$, and densities such as $\Obhtwo$ or $\Ocdmhtwo$.\footnote{The shift parameter $R$ controls the angular position of the acoustic peaks given fixed matter physics, and $\la$ encodes the angular size of the sound horizon at last scattering. Both are reviewed in Sec.~\ref{sec:framework}.} A Gaussian likelihood built from such a vector is fast, transparent, and easily embedded in any standard Boltzmann-code-plus-Markov-chain-Monte-Carlo (MCMC) pipeline. It is especially valuable for late-time extensions of the standard cosmological model, where the CMB primarily acts as a high-redshift geometric anchor through the acoustic scale and the physical densities~\cite{Bansal:2025xxx}.

This paper has two goals. First, we construct five model-adapted compressed CMB likelihoods---two for the dark-energy sector, two for extensions involving massive neutrinos, and one for spatial curvature---providing the data vectors and covariance matrices in a form that can be directly hard-coded into an inference pipeline. Second, we validate each compression against its corresponding reference chain by combining it with the DESI DR2 BAO likelihood and comparing the resulting marginalized constraints.

The CMB data underlying our compression are drawn from the most recent generation of small- and large-scale measurements: SPT-3G D1, ACT~DR6, \Planck\ PR3 primary CMB anisotropies, and \Planck\ PR4/NPIPE CMB lensing~\cite{SPT-3G:2022hvq,ACT:2023kun,Planck_PR3}. Throughout, we collectively denote this combined dataset by SPA. The mean vectors and covariance matrices of every compression are extracted from SPA chains, providing a single, internally consistent reference.

The paper is organized as follows. Section~\ref{sec:methodology} describes the cosmological models, the compression framework, and the inference setup. Section~\ref{sec:results} presents the model-by-model validation of the compressed likelihoods together with a comparison of marginalized constraints. We summarize and conclude in Sec.~\ref{sec:conclusion}. The full set of mean vectors and covariance matrices is collected in a single reference table (Table~\ref{tab:dvs_and_cov}) at the end of Sec.~\ref{sec:compression}.

% ============================================================
\section{Methodology, Models, and Data}
\label{sec:methodology}
% ============================================================

We describe in turn the cosmological models analyzed, the compression strategy, the inference pipeline, and the reference data. Throughout this work we assume general relativity. We impose spatial flatness in all models except \olcdm, where curvature is a free parameter.

\subsection{Background expansion}
\label{sec:background}

For all five model classes the dimensionless expansion rate takes the general form
\begin{equation}
\begin{split}
    E^2(z) \equiv \frac{H^2(z)}{H_0^2}
    &= \Om(1+z)^3 + \Omega_r(1+z)^4 \\
    &\quad + \Ok(1+z)^2 + \ode\,f_{\mathrm{de}}(z),
\end{split}
\label{eq:Ez_general}
\end{equation}
where $H_0$ is the present-day Hubble constant and $\Om$, $\Omega_r$, $\Ok$, and $\ode$ are the present-day fractional energy densities in matter, radiation, curvature, and dark energy, respectively.\footnote{We use the notational convention $h \equiv H_0/(100\kmsMpc)$ throughout. Physical (rather than fractional) densities are denoted by the conventional combinations $\Obhtwo$, $\Ocdmhtwo$, etc.} The function $f_{\mathrm{de}}(z)$ encodes the redshift dependence of the dark-energy component and reduces to unity for a cosmological constant.

The basic free parameters are $H_0$, the physical baryon density $\Obhtwo$, and $\Om$; specific models add further parameters as described in Sec.~\ref{sec:models}. In the models with variable neutrino mass, the neutrino energy density is evolved relativistically by CAMB and is therefore not treated as a pressureless matter component at recombination. At late times, after the massive-neutrino component has become non-relativistic, the present-day matter budget implies the derived cold-dark-matter density 
\begin{equation} 
\Ocdmhtwo = \Om h^2 - \frac{\sumnu}{93.14\eV} - \Obhtwo, \label{eq:omch2_derived} 
\end{equation} 
where $\sumnu$ is the sum of the neutrino masses. The factor $93.14\eV$ converts the summed neutrino mass into the present-day physical neutrino-density contribution, assuming the standard thermal history. Equation~\eqref{eq:omch2_derived} should therefore be understood as a late-time density-budget relation, not as an assumption that massive neutrinos behave as cold matter at recombination. Throughout, the effective number of relativistic species is fixed to $\Neff = 3.044$, and $\sumnu$ is fixed to its minimum-mass value $0.06\eV$ except in \nulcdm\ and \nuwowacdm, where it is varied.

\begin{table*}[tbp]
\centering
\caption{Summary of the cosmological models, sampled cosmological parameters, fixed parameters, and the corresponding compressed CMB likelihoods. The combined CMB dataset SPA refers to SPT-3G D1\,+\,ACT~DR6\,+\,\Planck\ PR3 primary CMB anisotropies\,+\,\Planck\ PR4/NPIPE CMB lensing. All compressions are derived from chains using SPA.}
\label{tab:model_summary}
\small
\setlength{\tabcolsep}{8pt}
\begin{tabular}{@{}llll@{}}
\toprule
Model & Sampled parameters & Fixed parameters & Compressed likelihood \\
\midrule
\lcdm     & $H_0,\;\Obhtwo,\;\Om$
          & $\sumnu = 0.06\eV$, $\Neff = 3.044$
          & \cmbthree:\;\; $(R,\,\la,\,\Obhtwo)$ \\[3pt]
\wowacdm  & $H_0,\;\Obhtwo,\;\Om,\;w_0,\;w_a$
          & $\sumnu = 0.06\eV$, $\Neff = 3.044$
          & \cmbthreew:\;\; $(R,\,\la,\,\Obhtwo)$ \\[3pt]
\nulcdm   & $H_0,\;\Obhtwo,\;\Om,\;\sumnu$
          & $\Neff = 3.044$
          & \cmbfournu:\; $(R,\,\la,\,\Obhtwo,\,\Ocdmhtwo)$ \\[3pt]
\olcdm    & $H_0,\;\Obhtwo,\;\Om,\;\Ok$
          & $\sumnu = 0.06\eV$, $\Neff = 3.044$
          & \cmbfourk:\;\; $(R,\,\la,\,\Obhtwo,\,\OL)$ \\[3pt]
% --- NEW ROW ---
\nuwowacdm & $H_0,\;\Obhtwo,\;\Om,\;\sumnu,\;w_0,\;w_a$
          & $\Neff = 3.044$
          & \cmbfournuw:\; $(R,\,\la,\,\Obhtwo,\,\Ocdmhtwo)$ \\
% ---------------
\bottomrule
\end{tabular}
\end{table*}

\subsection{Cosmological models}
\label{sec:models}

We consider five model classes, listed below in the order in which they appear in the validation. The models are paired with the compressed likelihoods according to which combinations of compressed observables are most sensitive to the relevant physics. We adopt the following naming convention for the compressed likelihoods:
\begin{itemize}\setlength{\itemsep}{2pt}
  \item \cmbthree: the three-parameter compression $(R,\,\la,\,\Obhtwo)$, applied to \lcdm.
  \item \cmbthreew: the three-parameter compression $(R,\,\la,\,\Obhtwo)$, applied to \wowacdm. It uses the same observable vector as \cmbthree\ but with mean vector and covariance extracted from \wowacdm\ SPA chains.
  \item \cmbfournu: the four-parameter neutrino compression $(R,\,\la,\,\Obhtwo,\,\Ocdmhtwo)$, applied to \nulcdm.
  \item \cmbfournuw: the four-parameter neutrino--dark-energy compression $(R,\,\la,\,\Obhtwo,\,\Ocdmhtwo)$, applied to \nuwowacdm. It uses the same observable vector as \cmbfournu\ but with mean vector and covariance re-extracted from \nuwowacdm\ SPA chains.
  \item \cmbfourk: the four-parameter curvature compression $(R,\,\la,\,\Obhtwo,\,\OL)$, applied to \olcdm.
\end{itemize}
We now specify parameterizations for the cosmological models that we consider. 

\medskip
\noindent 
\textbf{\lcdm.}
In the spatially flat cosmological-constant model, dark energy is a cosmological constant ($f_{\mathrm{de}}=1$, $\Ok=0$), so that
\begin{equation}
    E^2(z) = \Om(1+z)^3 + \Omega_r(1+z)^4 + (1 - \Om - \Omega_r).
\label{eq:Ez_lcdm}
\end{equation}
We sample the parameters $\{H_0,\,\Obhtwo,\,\Om\}$, with $\sumnu = 0.06\eV$, and use the \cmbthree\ compressed likelihood.

\medskip
\noindent 
\textbf{\wowacdm.}
For the Chevallier--Polarski--Linder (CPL) parameterization~\cite{Chevallier:2000qy,Linder:2002et}, the dark-energy equation of state varies linearly with scale factor as $w(a) = w_0 + w_a (1-a)$. This induces a dark-energy density evolution
\begin{equation}
  f_{\mathrm{de}}(z) = (1+z)^{3(1+w_0+w_a)}
  \exp\!\left[-\frac{3 w_a z}{1+z}\right].
\label{eq:fde_cpl}
\end{equation}
Assuming flatness  ($\Ok = 0$),  the expansion history follows Eq.~\eqref{eq:Ez_general} with $f_{\mathrm{de}}$ given above. The sampled parameters are $\{H_0,\,\Obhtwo,\,\Om,\,w_0,\,w_a\}$, with $\sumnu = 0.06\eV$. The resulting compressed likelihood is \cmbthreew, which uses the same observable vector $(R,\,\la,\,\Obhtwo)$ as \cmbthree\ but with mean vector and covariance re-extracted from \wowacdm\ SPA chains. No additional early-Universe density variable is required because the CPL parameters modify only the late-time expansion history: they change the distance to last scattering through the background evolution, but do not alter recombination physics, the sound-horizon calculation, or the primary-CMB physical-density constraints.

\medskip
\noindent 
\textbf{\nulcdm.}
The \nulcdm\ model keeps the flat \lcdm\ expansion history but promotes the summed neutrino mass to a free parameter, giving the parameter set $\{H_0,\,\Obhtwo,\,\Om,\,\sumnu\}$. The compressed likelihood is \cmbfournu, which augments the three geometric observables of \cmbthree\ with the cold-dark-matter density $\Ocdmhtwo$. This addition is necessary because varying $\sumnu$ changes the late-time matter budget through Eq.~\eqref{eq:omch2_derived}, while the primary CMB remains directly sensitive to the baryon and cold-dark-matter physical densities that determine the pre-recombination acoustic physics. A four-vector that explicitly tracks $\Ocdmhtwo$ is therefore required to retain the relevant neutrino-mass information without incorrectly treating massive neutrinos as cold matter at recombination.

\medskip
\noindent 
\textbf{\olcdm.}
Allowing curvature in \lcdm\ introduces an additional degree of freedom in the expansion history
\begin{equation}
\begin{split}
    E^2(z) &= \Om(1+z)^3 + \Omega_r(1+z)^4 + \Ok(1+z)^2 \\[0.1cm]
    &\quad + (1 - \Om - \Ok - \Omega_r).
\end{split}
\label{eq:Ez_oklcdm}
\end{equation}
The free parameters are $\{H_0,\,\Obhtwo,\,\Om,\,\Ok\}$, with $\sumnu = 0.06\eV$. The compressed likelihood is \cmbfourk, which augments the three geometric observables with the dark-energy density $\OL = 1 - \Om - \Ok - \Omega_r$. As discussed in Sec.~\ref{sec:compression}, in this case we additionally allow the redshift of last scattering $z_*$ to be evaluated self-consistently from CAMB rather than fixing it.

% --- NEW MODEL ---
\medskip
\noindent 
\textbf{\nuwowacdm.}
The \nuwowacdm\ model combines the CPL dark-energy parameterisation of \wowacdm\ with free neutrino mass, giving the parameter set $\{H_0,\,\Obhtwo,\,\Om,\,\sumnu,\,w_0,\,w_a\}$. The expansion history follows Eq.~\eqref{eq:Ez_general} with $f_{\mathrm{de}}$ given by Eq.~\eqref{eq:fde_cpl} and $\Ok = 0$. The compressed likelihood is \cmbfournuw, which uses the same observable vector $(R,\,\la,\,\Obhtwo,\,\Ocdmhtwo)$ as \cmbfournu\ but is distinct from it: the mean vector and covariance are extracted from dedicated \nuwowacdm\ SPA chains, since the enlarged parameter space shifts the posterior in the compressed-observable plane relative to \nulcdm. The \cmbfournuw\ notation is therefore reserved for the \nuwowacdm\ model throughout this paper; the corresponding data vector and covariance are given separately in Table~\ref{tab:dvs_and_cov}.
% -----------------

A summary of the five cosmological models, their free parameters that are sampled, and the resulting compressed likelihood are given in Table~\ref{tab:model_summary}.

\begin{table*}[tbp]
\centering
\caption{Mean vectors $\bar{\mathbf{v}}$ and covariance matrices $\mathbf{C}$ for the compressed CMB likelihoods, all extracted from SPA chains. \cmbthree\ uses $(R,\,\la,\,\Obhtwo)$ with $z_* = 1090$ and is used for \lcdm. \cmbthreew\ uses the same observable vector with $z_* = 1090$ but with mean and covariance extracted from \wowacdm\ chains; it is used for \wowacdm. \cmbfournu\ adds $\Ocdmhtwo$ to the geometric parameters, with $z_* = 1090$, and is used for \nulcdm. \cmbfourk\ adds $\OL$, with $z_*$ evaluated from CAMB at each MCMC step, and is used for \olcdm. \cmbfournuw\ uses the same observable vector as \cmbfournu\ but with mean and covariance extracted from \nuwowacdm\ chains; it is used exclusively for \nuwowacdm. Numbers are rounded to five significant figures for display. Together with Eq.~\eqref{eq:chi2comp}, the entries in this table define each compressed likelihood.}
\label{tab:dvs_and_cov}

\renewcommand{\arraystretch}{1.4}
\setlength{\tabcolsep}{3pt}
\setlength{\arraycolsep}{3pt}
\footnotesize

\newsavebox{\dvstable}
\begin{lrbox}{\dvstable}
\begin{tabular}{@{} c c c c @{}}
\toprule
\begin{tabular}{@{}c@{}}Compression\\[-0.15cm](model)\end{tabular}
&
Mean vector $\bar{\mathbf{v}}$
&
Covariance matrix $\mathbf{C}$
&
$z_*$ \\
\midrule

\begin{tabular}{@{}c@{}}\cmbthree\\[-0.15cm](\lcdm)\end{tabular}
&
$\begin{pmatrix} R \\ \la \\ \Obhtwo \end{pmatrix}
=
\begin{pmatrix} 1.7537 \\ 301.82 \\ 0.022381 \end{pmatrix}$
&
$\begin{pmatrix}
    9.8079 \times 10^{-6} & -1.3268 \times 10^{-6} & -1.2316 \times 10^{-7} \\
   -1.3268 \times 10^{-6} &  4.8132 \times 10^{-3} &  1.7615 \times 10^{-6} \\
   -1.2316 \times 10^{-7} &  1.7615 \times 10^{-6} &  8.8429 \times 10^{-9}
\end{pmatrix}$
& $1090$ \\[4ex]

\midrule

\begin{tabular}{@{}c@{}}\cmbthreew\\[-0.15cm](\wowacdm)\end{tabular}
&
$\begin{pmatrix} R \\ \la \\ \Obhtwo \end{pmatrix}
=
\begin{pmatrix} 1.7481 \\ 301.84 \\ 0.022451 \end{pmatrix}$
&
$\begin{pmatrix}
    1.0213 \times 10^{-5} & -3.0832 \times 10^{-6} & -1.3030 \times 10^{-7} \\
   -3.0832 \times 10^{-6} &  4.9891 \times 10^{-3} &  1.8345 \times 10^{-6} \\
   -1.3030 \times 10^{-7} &  1.8345 \times 10^{-6} &  9.2580 \times 10^{-9}
\end{pmatrix}$
& $1090$ \\[4ex]

\midrule

\begin{tabular}{@{}c@{}}\cmbfournu\\[-0.15cm](\nulcdm)\end{tabular}
&
$\begin{pmatrix} R \\ \la \\ \Obhtwo \\ \Ocdmhtwo \end{pmatrix}
=
\begin{pmatrix} 1.7528 \\ 301.82 \\ 0.022389 \\ 0.12097 \end{pmatrix}$
&
$\begin{pmatrix}
    2.3307 \times 10^{-5} & -2.8774 \times 10^{-5} & -1.7660 \times 10^{-7} &  4.0826 \times 10^{-6} \\
   -2.8774 \times 10^{-5} &  4.6149 \times 10^{-3} &  1.8150 \times 10^{-6} & -1.1680 \times 10^{-5} \\
   -1.7660 \times 10^{-7} &  1.8150 \times 10^{-6} &  9.1427 \times 10^{-9} & -3.5494 \times 10^{-8} \\
    4.0826 \times 10^{-6} & -1.1680 \times 10^{-5} & -3.5494 \times 10^{-8} &  1.0508 \times 10^{-6}
\end{pmatrix}$
& $1090$ \\[5ex]

\midrule

\begin{tabular}{@{}c@{}}\cmbfournuw\\[-0.15cm](\nuwowacdm)\end{tabular}
&
$\begin{pmatrix} R \\ \la \\ \Obhtwo \\ \Ocdmhtwo \end{pmatrix}
=
\begin{pmatrix} 1.7449 \\ 301.84 \\ 0.022436 \\ 0.11959 \end{pmatrix}$
&
$\begin{pmatrix}
    1.2694 \times 10^{-5} & -1.4712 \times 10^{-5} & -1.6062 \times 10^{-7} &  3.9537 \times 10^{-6} \\
   -1.4712 \times 10^{-5} &  4.7735 \times 10^{-3} &  1.9244 \times 10^{-6} & -1.1173 \times 10^{-5} \\
   -1.6062 \times 10^{-7} &  1.9244 \times 10^{-6} &  9.7018 \times 10^{-9} & -4.2755 \times 10^{-8} \\
    3.9537 \times 10^{-6} & -1.1173 \times 10^{-5} & -4.2755 \times 10^{-8} &  1.2558 \times 10^{-6}
\end{pmatrix}$
& $1090$ \\[5ex]

\midrule

\begin{tabular}{@{}c@{}}\cmbfourk\\[-0.15cm](\olcdm)\end{tabular}
&
$\begin{pmatrix} R \\ \la \\ \Obhtwo \\ \OL \end{pmatrix}
=
\begin{pmatrix} 1.7458 \\ 301.79 \\ 0.022480 \\ 0.66652 \end{pmatrix}$
&
$\begin{pmatrix}
    1.7393 \times 10^{-5} &  1.1390 \times 10^{-4} & -2.1009 \times 10^{-7} &  1.2378 \times 10^{-5} \\
    1.1390 \times 10^{-4} &  5.0689 \times 10^{-3} & -1.3944 \times 10^{-6} &  2.2002 \times 10^{-5} \\
   -2.1009 \times 10^{-7} & -1.3944 \times 10^{-6} &  1.0270 \times 10^{-8} & -1.2661 \times 10^{-7} \\
    1.2378 \times 10^{-5} &  2.2002 \times 10^{-5} & -1.2661 \times 10^{-7} &  1.4940 \times 10^{-4}
\end{pmatrix}$
& CAMB \\[3ex]

\bottomrule
\end{tabular}
\end{lrbox}
\colorbox{yellow!20}{\usebox{\dvstable}}
\end{table*}

\subsection{Compressed CMB likelihood}
\label{sec:compression}

\subsubsection{General framework}
\label{sec:framework}

The two core geometric observables in the compressed vector are the CMB shift parameter $R$ and the acoustic angular scale $\la$, defined as
\begin{equation}
  R \equiv \sqrt{\Om}\,H_0\frac{\DM(z_*)}{c},
  \qquad
  \la \equiv \pi\frac{\DM(z_*)}{r_s(z_*)},
\label{eq:Rla}
\end{equation}
where $\DM(z_*)$ is the comoving transverse distance to the last-scattering surface, $r_s(z_*)$ is the comoving sound horizon at the same epoch, and $z_*$ is the redshift of last scattering (set, depending on the model, either to a reference value or computed at each MCMC step). Physically, $R$ measures the angular position of the acoustic peaks for fixed matter physics, while $\la$ measures the angular size of the sound horizon. These two numbers carry the dominant information that the CMB provides about the expansion history. Depending on the model, this geometric core is augmented by one density parameter to form the full compressed data vector $\mathbf{v}$.

The compressed likelihood takes the Gaussian form
\begin{equation}
  -2 \ln \mathcal{L}_{\mathrm{comp}}
  = \bigl(\mathbf{v}_{\mathrm{th}} - \bar{\mathbf{v}}\bigr)^{\!\top}
    \mathbf{C}^{-1}
    \bigl(\mathbf{v}_{\mathrm{th}} - \bar{\mathbf{v}}\bigr),
\label{eq:chi2comp}
\end{equation}
where $\bar{\mathbf{v}}$ and $\mathbf{C}$ are the mean vector and covariance matrix obtained from a SPA chain in the relevant model and hard-coded into the Cobaya~\cite{Torrado:2020dgo} likelihood module. 
% \dragan{Comment on why C doesn't depend on cosmo parameters?} \amogh{Since the compressed geometric observables are extremely well-constrained by the CMB data, their localized posterior is highly Gaussian. Consequently, the covariance matrix $\mathbf{C}$ is determined by the experimental noise and dataset sensitivity, remaining effectively independent of the cosmological parameters across the viable region being sampled.} 
% \dragan{
The covariance matrix is treated as fixed because the compressed likelihood is constructed as a Gaussian approximation to the distribution of the compressed observables measured from the reference SPA chains. In this representation, the data product is the pair $(\bar{\mathbf{v}},\mathbf{C})$, while cosmology enters only through the theoretical prediction $\mathbf{v}_{\mathrm{th}}(\theta)$. A parameter-dependent covariance would be required only if the local width or orientation of the compressed-observable posterior changed appreciably across the sampled region. For the late-time extensions considered here, the CMB tightly constrains the compressed geometric variables, and the validation against full-CMB chains shows that a fixed covariance is sufficient over the posterior volume explored.
% }
The theoretical prediction $\mathbf{v}_{\mathrm{th}}$ is computed using CAMB~\cite{Lewis:1999bs} at each MCMC step.

For the three-parameter compressions \cmbthree\ and \cmbthreew\ and the four-parameter neutrino compressions \cmbfournu\ and \cmbfournuw, we fix the redshift of last scattering at $z_* = 1090$. This is an excellent approximation for flat, late-time-only extensions of \lcdm, because in these models $z_*$ depends on the matter and baryon densities through tightly constrained recombination physics that varies only weakly across the posterior. By contrast, in \olcdm\ the geometry of last scattering can shift more substantially because curvature alters the mapping between the sound horizon and its observed angular scale. We therefore allow $z_*$ to be evaluated self-consistently from CAMB at each MCMC step in the \cmbfourk\ pipeline, ensuring that the compressed likelihood remains accurate over the full curvature posterior.

\subsubsection{Data vectors and covariances}
\label{sec:dvs}

The mean vectors and covariance matrices of the compressions, all obtained from SPA chains in the relevant model, are collected in Table~\ref{tab:dvs_and_cov}. Each horizontal block corresponds to one compressed likelihood: \cmbthree\ (used in \lcdm), \cmbthreew\ (used in \wowacdm), \cmbfournu\ (used in \nulcdm), \cmbfourk\ (used in \olcdm), and \cmbfournuw\ (used in \nuwowacdm). For each compression, we list the mean of the compressed vector and the corresponding symmetric covariance matrix. These are the quantities that should be hard-coded into a user pipeline. Together with Eq.~\eqref{eq:chi2comp}, they fully define the compressed CMB likelihood.

\subsection{Inference pipeline and data}
\label{sec:pipeline}

We evaluate the compressed CMB likelihoods (\cmbthree, \cmbthreew, \cmbfournu, \cmbfournuw, \cmbfourk) within the Cobaya inference framework~\cite{Torrado:2020dgo}, using CAMB~\cite{Lewis:1999bs} to compute the comoving distance, the sound horizon, and associated derived quantities at each MCMC step.

Each compressed likelihood is combined with the DESI DR2 BAO likelihood (\path{bao.desi_dr2.desi_bao_all})~\cite{DESI:2025zgx}, which provides 13 distance measurements with their full covariance matrix and serves as the late-time geometric anchor. Posteriors are sampled with the Cobaya MCMC sampler,
using \texttt{covmat:auto}, \texttt{max\_tries=10000}, and the convergence criterion \texttt{Rminus1\_stop=0.01}. \footnote{For \wowacdm, the convergence criterion is modified to \texttt{Rminus1\_stop=0.001} to remove sampling artefacts in the posterior.}
% \prakhar{Should we change it to 0.001?}
Chains are post-processed and analyzed with GetDist~\cite{Lewis:2019xzd}.

For validation, we compare each compressed-plus-BAO posterior against the corresponding reference posterior obtained by combining DESI DR2 BAO with the full SPA CMB likelihood, comprising SPT-3G D1, ACT~DR6, \Planck\ PR3 primary CMB anisotropies, and \Planck\ PR4/NPIPE CMB lensing~\cite{SPT-3G:2022hvq,ACT:2023kun,Planck_PR3}. In the \Planck\ part of SPA, PR3 is used for the primary temperature and polarization anisotropy likelihoods, while PR4/NPIPE is used only for the CMB lensing-potential likelihood. Using SPA chains uniformly across all five model classes ensures that any residual differences between the compressed and reference posteriors arise from the compression itself rather than from a mismatch in CMB data choice.

\begin{figure*}[htbp]
  \centering
  % First Subfigure: Lambda CDM
  \subfloat[\lcdm\label{fig:lcdm}]{%
    \includegraphics[width=0.48\textwidth]{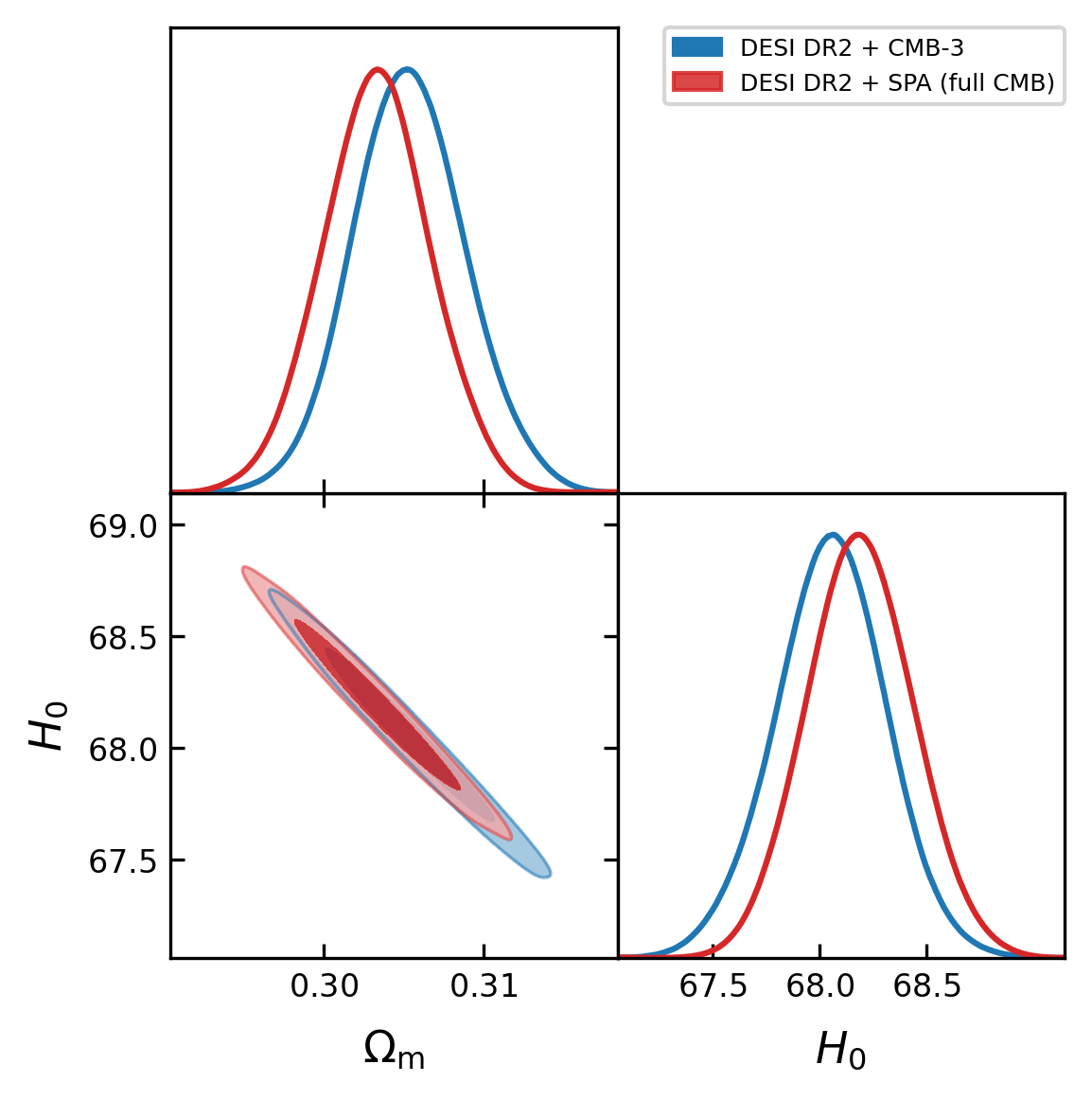}%
  }
  \hfill
  % Second Subfigure: w0waCDM
  \subfloat[\wowacdm\label{fig:w0wa}]{%
    \includegraphics[width=0.48\textwidth]{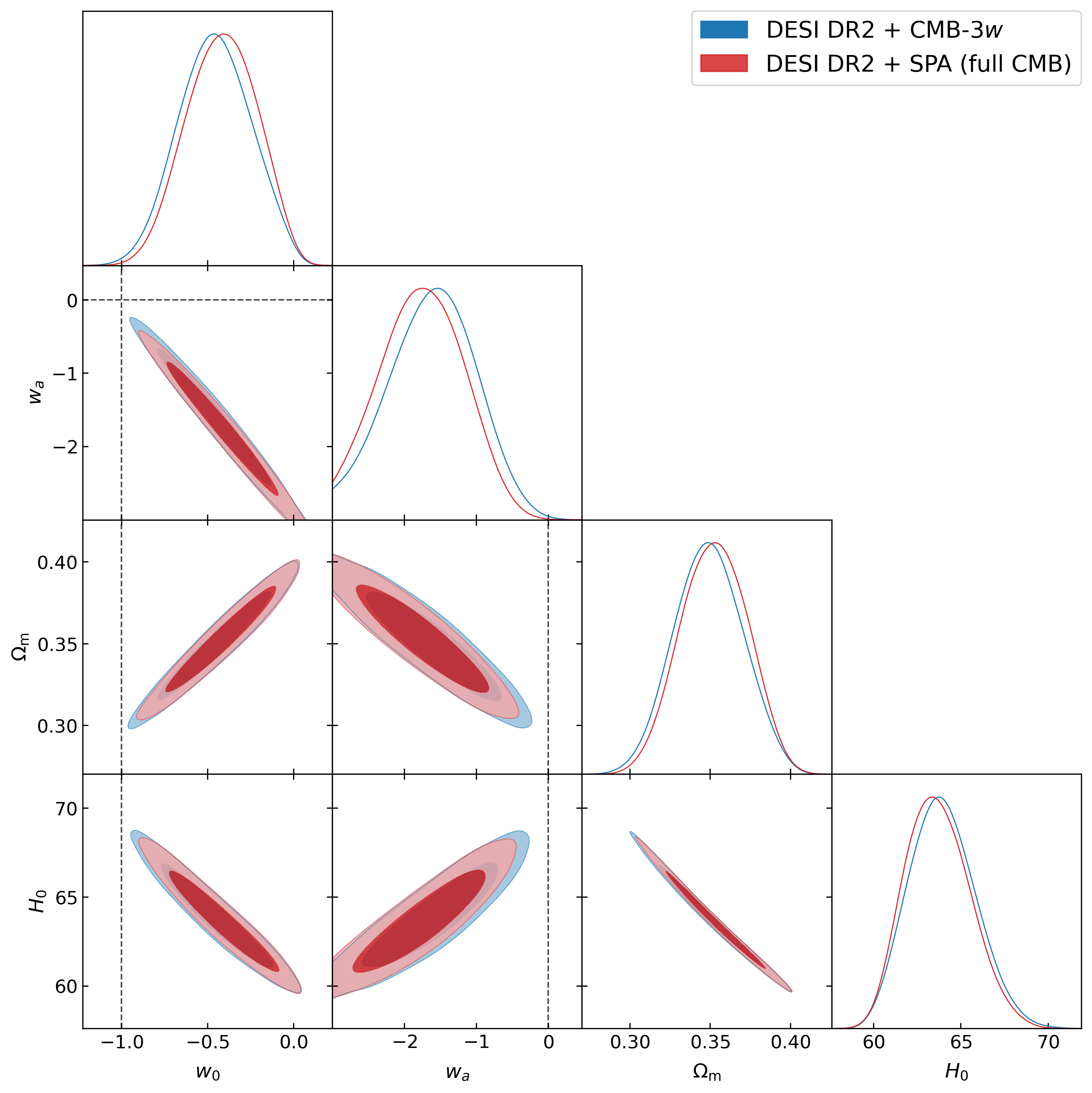}%
  }
  
  % Unified Main Caption
  \caption{Validation of the three-parameter compressions against SPA (full CMB, red) using DESI DR2 BAO in (a) \lcdm\ and (b) \wowacdm. \textbf{(a)} \lcdm\ constraints on $\Omega_m$ and $H_0$ from \cmbthree. \textbf{(b)} Full corner plot for the \wowacdm\ model from \cmbthreew. Dashed lines mark the \lcdm\ reference values $w_0 = -1$, $w_a = 0$. The 68\% and 95\% contours overlap closely across all parameters, demonstrating that \cmbthreew\ remains accurate in the extended dark energy parameter space.}
  \label{fig:combined_validation}
\end{figure*}

\vspace{-0.3cm}
\section{Results}
\label{sec:results}
% ============================================================

We now turn to the validation of the compressed likelihoods constructed in Sec.~\ref{sec:compression}. The principal results are presented in Figs.~\ref{fig:lcdm}--\ref{fig:nuwowacdm} and summarized in Table~\ref{tab:marginalized_constraints}. In every figure, blue contours show DESI DR2 BAO combined with the relevant compressed CMB likelihood, while red contours show the corresponding reference: DESI DR2 BAO combined with the full CMB likelihood. We discuss each model class in turn, in the same order as Table~\ref{tab:model_summary}.

\subsection{\lcdm}
\label{sec:lcdm}

Figure~\ref{fig:lcdm} shows the validation of \cmbthree\ in \lcdm. The compressed (DESI DR2 + \cmbthree) run yields
\begin{equation}
\begin{split}
  \Om &= 0.3054 \pm 0.0035, \\
  H_0 &= (68.05 \pm 0.26)\kmsMpc,
\end{split}
\end{equation}

while the DESI DR2 + SPA (full CMB) gives
\begin{equation}
\begin{split}
  \Om &= 0.3033 \pm 0.0034, \\
  H_0 &= (68.19 \pm 0.25)\kmsMpc.
\end{split}
\end{equation}
The compressed and reference posteriors trace the same $\Om$--$H_0$ degeneracy direction, and the 68\% and 95\% contours overlap to a high degree. The compressed posterior is, however, visibly shifted relative to the reference, as can be seen in Fig.~\ref{fig:lcdm}: the compressed mean of $H_0$ lies $\Delta H_0 \approx -0.14\kmsMpc$ below the reference value, while the compressed mean of $\Om$ lies $\Delta \Om \approx +0.0021$ above it. These shifts correspond to roughly $0.6\sigma$ in each case, which is below the level at which they would affect any current or near-term cosmological inference. 

% \dragan{heavily redacted from here on:} 
These modest but non-negligible biases in $\Om$ and $H_0$ in \lcdm\ are likely due to the gravitational lensing of the CMB which is not explicitly modeled in the data compression. The inclusion of CMB lensing preferentially shifts the Hubble constant downwards (see e.g.\ \cite{SPT-3G:2025bzu}) and hence $\Om$ upwards. In a cosmological models with more degrees of freedom, such as \wowacdm\ and its variants that we discuss next, the additional dark energy parameters can take values to match the way that CMB lensing (and perhaps other not-explicitly-modeled effects, such as the late-time Integrated Sachs-Wolfe effect) are captured by the compressed parameters, leading to essentially no bias in the compression-derived cosmological parameters. In contrast, \lcdm\ does not have sufficiently many degrees of freedom to make such a matching  possible, leaving small residual biases in the cosmological parameters.

\subsection{\wowacdm}
\label{sec:w0wa}
% \dragan{???? Did We ever shift to w0waCDM? Missing title? I dont see how there are 5 pars in LCDM.}

Figure~\ref{fig:w0wa} shows \cmbthreew\ applied in the enlarged CPL parameter space. The compressed chain (DESI DR2 + \cmbthreew) yields
\begin{equation}
\begin{split}
  w_0 &= -0.46 \pm 0.21, \\
  w_a &= -1.60 \pm 0.59, \\
  \Om &= 0.3491 \pm 0.0212, \\
  H_0 &= (63.95 \pm 1.89)\kmsMpc,
\end{split}
\end{equation}
while the DESI DR2 + SPA (full CMB) reference gives
\begin{equation}
\begin{split}
  w_0 &= -0.42 \pm 0.20, \\
  w_a &= -1.74 \pm 0.57, \\
  \Om &= 0.3526 \pm 0.0207, \\
  H_0 &= (63.70 \pm 1.84)\kmsMpc.
\end{split}
\end{equation}

Even in this five-parameter CPL extension, \cmbthreew\ reproduces the posterior geometry with good accuracy. The contours in the $(w_0, w_a)$ plane track the reference closely, and the compressed posteriors of $\Om$ and $H_0$ broaden in the expected way relative to \lcdm\ as the dark-energy parameters open up additional degeneracy directions.

% As in \lcdm, the agreement is not exact \dragan{I disagree with this sentence; the match is much better than in \lcdm, while "nothing is perfect" is always true of course}. REWYUIIP
The agreement between the compressed and full-CMB constraints is excellent. The compressed means of $w_0$ and $w_a$ are shifted slightly relative to the full-CMB reference (by about $0.2\sigma$ and $0.3\sigma$, respectively), with similarly small offsets in $\Om$ and $H_0$. These residuals reflect the same underlying limitation discussed in Sec.~\ref{sec:lcdm}: the compressed likelihood is a Gaussian summary of a few geometric observables and therefore does not inherently reproduce the full CMB posterior, so a small bias is expected even when, as here, the compression is extracted from chains of the same model. In this case, however, the offsets are absorbed within the much wider $w_0$--$w_a$ contours and play no perceptible role in the validation. The result is nonetheless non-trivial: it shows that a three-observable compression retains its accuracy across a substantial extension of the late-time expansion history, where the dark-energy parameters open up degeneracy directions far broader than those present in \lcdm.

%\vspace{-1cm}

\begin{figure*}[htbp]
  \centering
  \subfloat[\nulcdm\label{fig:nulcdm}]{%
    \includegraphics[width=0.48\textwidth]{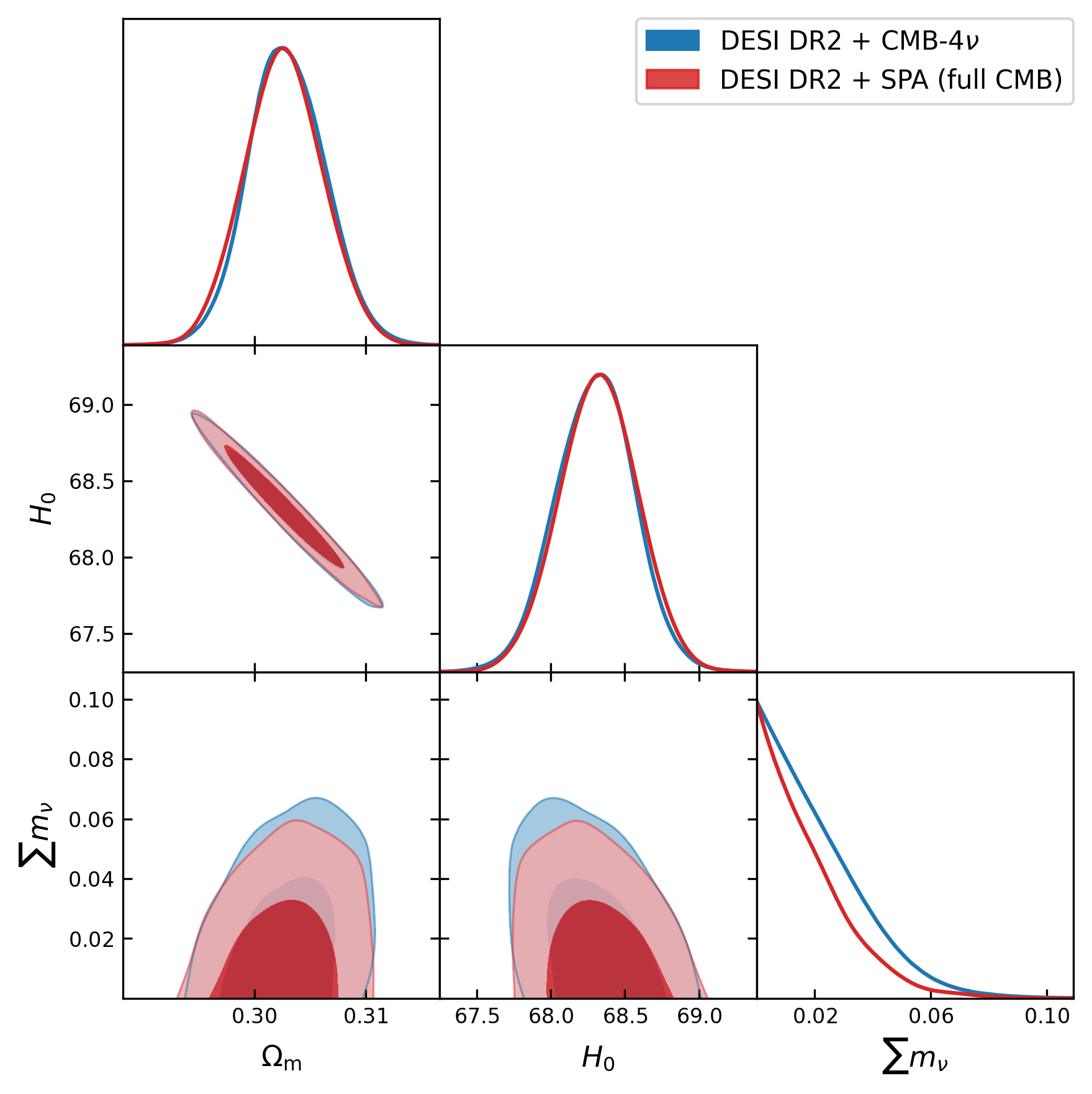}%
  }
  \hfill
  \subfloat[\olcdm\label{fig:olcdm}]{%
    \includegraphics[width=0.48\textwidth]{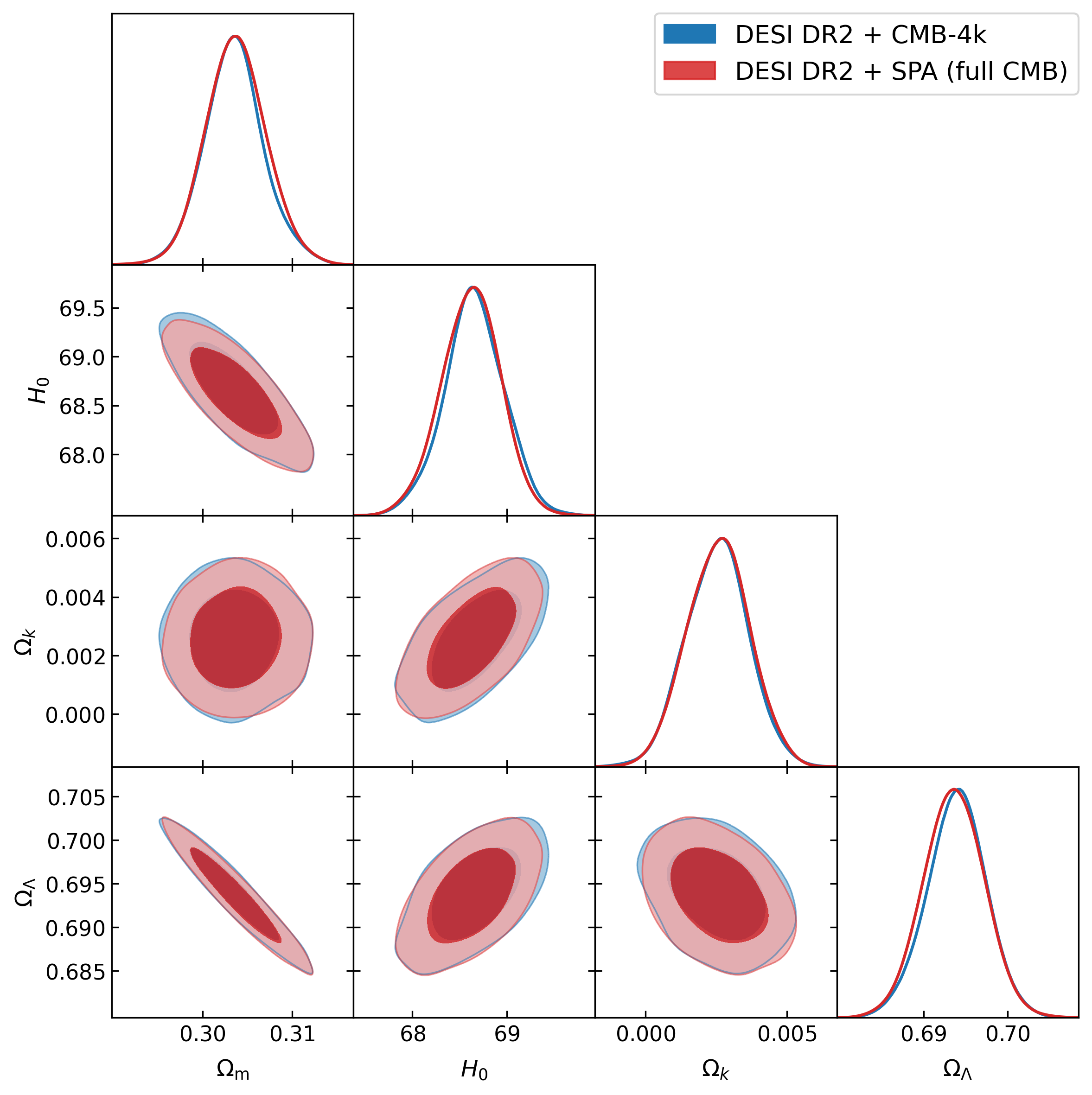}%
  }
  
  \caption{Validation of compressed (blue) and full (red) CMB likelihoods. \textbf{(a)} \nulcdm: DESI DR2 BAO + \cmbfournu\ (blue) vs.\ DESI DR2 BAO + SPA (full CMB) (red). Agreement is excellent for $\Omega_m$ and $H_0$, and very good for the one-sided neutrino mass posterior, with the compressed upper limit slightly looser than the reference one. \textbf{(b)} \olcdm: DESI DR2 BAO + \cmbfourk\ (blue) vs.\ DESI DR2 BAO + SPA (full CMB) (red). The compressed posterior tracks $\Omega_m$, $H_0$, $\Omega_k$, and $\Omega_\Lambda$ with high fidelity; residual shifts are below the level of $0.1\sigma$ for every parameter.}
  \label{fig:combined_validation_extensions}
\end{figure*}

\subsection{\nulcdm}
\label{sec:nulcdm}
\vspace{-0.2cm}
Figure~\ref{fig:nulcdm} shows the validation of \cmbfournu\ in \nulcdm. The compressed run (DESI DR2 + \cmbfournu) yields
\begin{equation}
\begin{split}
  \Om &= 0.3030 \pm 0.0034, \\
  H_0 &= (68.30 \pm 0.26)\kmsMpc, \\
  \sumnu &< 0.053\eV \quad (95\%),
\end{split}
\end{equation}
to be compared with the DESI DR2 + SPA (full CMB) reference:
\vspace{-1em}
\begin{equation}
\begin{split}
  \Om &= 0.3026 \pm 0.0035, \\
  H_0 &= (68.32 \pm 0.26)\kmsMpc, \\
  \sumnu &< 0.047\eV \quad (95\%).
\end{split}
\end{equation}

%\vspace{-1.0em}

The $\Om$ and $H_0$ posteriors are reproduced with excellent accuracy, with mean offsets well below $0.2\sigma$. The neutrino-mass posterior is also recovered well: the compressed upper limit is somewhat looser than the reference one ($0.053\eV$ vs.\ $0.047\eV$), reflecting a slightly broader high-mass tail in the compressed chain. This is the expected behavior of \cmbfournu, since $\sumnu$ enters the compressed likelihood only indirectly through the matter-budget relation~\eqref{eq:omch2_derived}: the compression carries less leverage on $\sumnu$ than the full CMB likelihood, but enough to deliver a faithful one-sided posterior at the $\sim 10\%$ level on the upper limit. 

\subsection{\olcdm}
\label{sec:olcdm}
Figure~\ref{fig:olcdm} shows the validation of \cmbfourk\ in \olcdm. The compressed run (DESI DR2 + \cmbfourk) yields
\vspace{-0.5em}
\begin{equation}
\begin{split}
  \Om &= 0.3035 \pm 0.0033, \\
  H_0 &= (68.66 \pm 0.32)\kmsMpc, \\
  \Ok &= 0.0025 \pm 0.0011, \\
  \OL &= 0.6939 \pm 0.0035,
\end{split}
\end{equation}
and the DESI DR2 + SPA (full CMB) chains give
\begin{equation}
\begin{split}
  \Om &= 0.3037 \pm 0.0034, \\
  H_0 &= (68.62 \pm 0.32)\kmsMpc, \\
  \Ok &= 0.0026 \pm 0.0011, \\
  \OL &= 0.6936 \pm 0.0036.
\end{split}
\end{equation}
The \cmbfourk\ compression captures all four parameters with high fidelity: the means agree to better than $0.1\sigma$ in every case, and the contour shapes and degeneracy directions in the $(\Om, H_0, \Ok, \OL)$ subspace are essentially indistinguishable between blue and red. The fact that the residual offsets seen in the \lcdm\ case (Fig.~\ref{fig:lcdm}) are essentially absent here illustrates the value of (i) augmenting the compression with a fourth observable that is naturally suited to the model, and (ii) evaluating $z_*$ self-consistently from CAMB rather than fixing it. Together, these two refinements remove the small geometric leakage that \cmbthree\ exhibits in \lcdm.

\begin{figure}[htbp]
  \centering
  \includegraphics[width=0.5\textwidth]{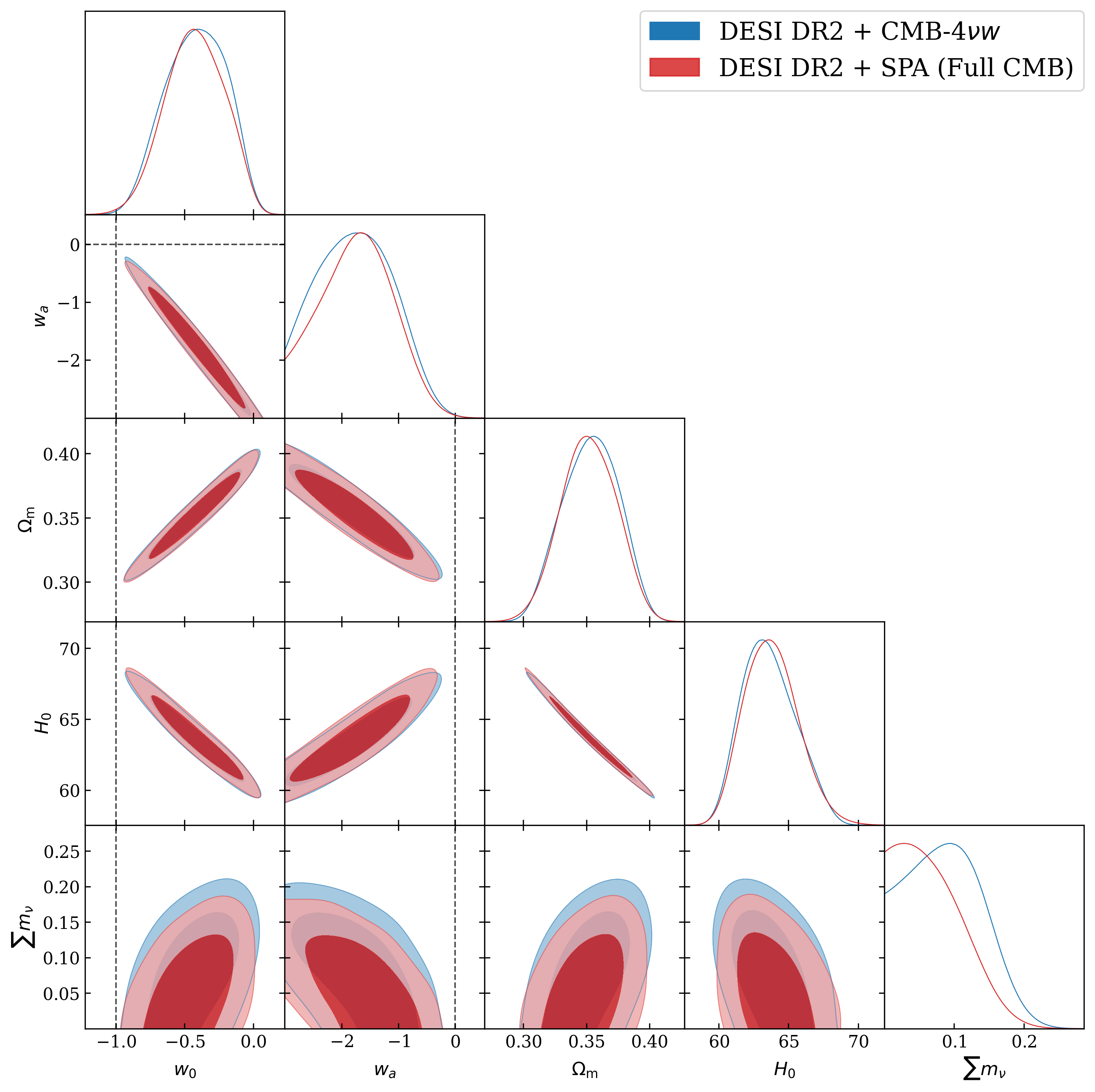}
  \caption{Validation of \cmbfournuw\ in \nuwowacdm: DESI DR2 BAO + \cmbfournuw\ (blue) vs.\ DESI DR2 BAO + SPA (full CMB) (red). The compressed posterior reproduces all five parameters $(w_0,\,w_a,\,\Om,\,H_0,\,\sumnu)$ with good accuracy. Dashed lines mark the \lcdm\ reference values $w_0=-1$, $w_a=0$. The neutrino-mass constraint is somewhat looser in the compressed run than in the reference, consistent with the reduced leverage of the four-parameter compression on $\sumnu$ when dark-energy degrees of freedom are simultaneously varied.}
  \label{fig:nuwowacdm}
\end{figure}

% --- NEW SUBSECTION ---
\subsection{\nuwowacdm}
\label{sec:nuwowacdm}

Figure~\ref{fig:nuwowacdm} shows the validation of \cmbfournuw\ in the joint \nuwowacdm\ model. The compressed run (DESI DR2 + \cmbfournuw) yields
\begin{equation}
\begin{split}
  w_0 &= -0.416 \pm 0.219, \\
  w_a &= -1.738 \pm 0.649, \\
  \Om &= 0.3536 \pm 0.0221, \\
  H_0 &= (63.61 \pm 1.91)\kmsMpc, \\
  \sumnu &= (0.090 \pm 0.051)\eV,
\end{split}
\end{equation}
while the DESI DR2 + SPA (full CMB) reference gives
\begin{equation}
\begin{split}
  w_0 &= -0.424 \pm 0.213, \\
  w_a &= -1.727 \pm 0.621, \\
  \Om &= 0.3520 \pm 0.0217, \\
  H_0 &= (63.77 \pm 1.91)\kmsMpc, \\
  \sumnu &= (0.070 \pm 0.046)\eV.
\end{split}
\end{equation}

The compressed posterior reproduces all five parameters with good accuracy. The dark-energy parameters $(w_0,\,w_a)$ are recovered to within $0.3\sigma$, consistent with the performance of \cmbthreew\ in \wowacdm\ (Sec.~\ref{sec:w0wa}). The $\Om$ and $H_0$ contours are likewise well matched. The neutrino-mass posterior peaks away from zero in both the compressed and reference chains, in contrast to the upper-limit behaviour seen in \nulcdm; the compressed mean is somewhat higher than the reference ($0.090\eV$ vs.\ $0.070\eV$), while the $95\%$ upper limits are $0.175\eV$ and $0.153\eV$, respectively. This modest offset is consistent with the reduced leverage of \cmbfournuw\ on $\sumnu$ noted in Sec.~\ref{sec:nulcdm}: the compression tracks $\sumnu$ only through the matter-budget relation~\eqref{eq:omch2_derived}, and the additional degeneracy directions opened by $(w_0,\,w_a)$ slightly dilute the constraining power on neutrino mass. Nonetheless, the qualitative structure of the neutrino-mass posterior including its peak location and width is faithfully reproduced.

\begin{table*}[htbp]
\centering
\caption{Marginalized constraints from the compressed CMB\,+ DESI runs and the corresponding reference DESI\,+ SPA (full CMB). For each parameter, we list the mean and 68\% uncertainty. Upper limits on $\sumnu$ are one-tailed 95\% bounds (indicated by $<$); two-sided 68\% constraints are listed otherwise. Models are listed in the same order as Table~\ref{tab:model_summary}.}
% \dragan{[need fewer digits in H0 in w0waCDM.]}}
\label{tab:marginalized_constraints}
\renewcommand{\arraystretch}{1.3} % Vertical spacing
\begin{tabular}{
  ll 
  % A single S column per data block handling the \pm natively
  S[table-format=-2.4 \pm 1.4, separate-uncertainty=true] 
  @{\hspace{3em}} 
  S[table-format=-2.4 \pm 1.4, separate-uncertainty=true]
}
\toprule
Model & Parameter & {Compressed: 68\%} & {Full CMB: 68\%} \\
\midrule
\multirow{2}{*}{\lcdm\ (\cmbthree)}
  & $\Om$                 & 0.3054 \pm 0.0035 & 0.3033 \pm 0.0034 \\
  & $H_0\;[\kmsMpc]$      & 68.05 \pm 0.26    & 68.19 \pm 0.25 \\
\midrule
\multirow{4}{*}{\wowacdm\ (\cmbthreew)}
  & $w_0$                 & -0.4559 \pm 0.2087 & -0.4161 \pm 0.2028 \\
  & $w_a$                 & -1.6049 \pm 0.5891 & -1.7426 \pm 0.5666 \\
  & $\Om$                 & 0.3491 \pm 0.0212  & 0.3526 \pm 0.0207 \\
  & $H_0\;[\kmsMpc]$      & 63.95 \pm 1.89 & 63.70 \pm 1.84 \\
\midrule
\multirow{3}{*}{\nulcdm\ (\cmbfournu)}
  & $\Om$                 & 0.3030 \pm 0.0034  & 0.3026 \pm 0.0035 \\
  & $H_0\;[\kmsMpc]$      & 68.30 \pm 0.26     & 68.32 \pm 0.26 \\
  % Enclosing non-numeric data in braces {} escapes the S-column parser and centers it
  & $\sumnu\;[\eV]$       & {$<0.053$ (95\%)}  & {$<0.047$ (95\%)} \\
\midrule
\multirow{4}{*}{\olcdm\ (\cmbfourk)}
  & $\Om$                 & 0.3035 \pm 0.0033  & 0.3037 \pm 0.0034 \\
  & $H_0\;[\kmsMpc]$      & 68.66 \pm 0.32     & 68.62 \pm 0.32 \\
  & $\Ok$                 & 0.0025 \pm 0.0011  & 0.0026 \pm 0.0011 \\
  & $\OL$                 & 0.6939 \pm 0.0035  & 0.6936 \pm 0.0036 \\
\midrule
\multirow{5}{*}{\nuwowacdm\ (\cmbfournuw)}
  & $w_0$                 & -0.416 \pm 0.219   & -0.424 \pm 0.213 \\
  & $w_a$                 & -1.738 \pm 0.649   & -1.727 \pm 0.621 \\
  & $\Om$                 & 0.3536 \pm 0.0221  & 0.3520 \pm 0.0217 \\
  & $H_0\;[\kmsMpc]$      & 63.61 \pm 1.91     & 63.77 \pm 1.91 \\
  & $\sumnu\;[\eV]$       & 0.090 \pm 0.051    & 0.070 \pm 0.046 \\
\bottomrule
\end{tabular}
\end{table*}

\subsection{Marginalized constraint comparison}
\label{sec:table}

Table~\ref{tab:marginalized_constraints} collects the 68\% marginalized means and the 95\% central intervals for every parameter shown in Figs.~\ref{fig:lcdm}--\ref{fig:nuwowacdm}. Across the five model classes, the compressed likelihoods reproduce the reference posteriors at a level suitable for practical late-time cosmological inference. 
% \dragan{say what fraction here as well}
The agreement is essentially exact in \olcdm, where all listed mean shifts are below about $0.13\sigma$. In \wowacdm, all mean shifts are below about $0.3\sigma$. In \nulcdm, the shifts in $\Om$ and $H_0$ are below $0.12\sigma$, while the one-sided neutrino-mass upper limit is only about $13\%$ weaker than the full-CMB reference. In \nuwowacdm, the dark-energy and geometric parameters agree within $0.09\sigma$, while the mean value of $\sumnu$ is shifted by about $0.4\sigma$ and its 95\% upper interval is about $14\%$ broader. The only visibly larger residual in the minimal models is the \lcdm\ shift along the $\Om$--$H_0$ degeneracy, which remains below about $0.7\sigma$. The pattern is consistent with the structural differences between the compressions: where the compression includes a fourth observable that is well-matched to the model's physics (\nulcdm, \olcdm, \nuwowacdm), the agreement is closest, while the three-parameter compression in \lcdm\ exhibits the small residual shifts discussed in Sec.~\ref{sec:lcdm}.
% ============================================================
\section{Conclusions}
\label{sec:conclusion}
% ============================================================

We have presented \textsc{CMBComp}, a compressed CMB likelihood designed for fast late-time cosmological inference. The compression distills the dominant geometric information of the CMB into a small Gaussian likelihood built from the shift parameter $R$, the acoustic angular scale $\la$, and one or two physical density parameters. Five specific compressions are provided, all derived from the combined SPT-3G\,+\,ACT~DR6\,+\,\Planck\ NPIPE/PR4 (SPA) chains: \cmbthree, using $(R,\,\la,\,\Obhtwo)$ for \lcdm; \cmbthreew, using the same observable vector for \wowacdm\ but with mean and covariance re-extracted from \wowacdm\ SPA chains; \cmbfournu, using $(R,\,\la,\,\Obhtwo,\,\Ocdmhtwo)$ for \nulcdm; \cmbfournuw, using the same observable vector for \nuwowacdm\ but with mean and covariance re-extracted from \nuwowacdm\ SPA chains; and \cmbfourk, using $(R,\,\la,\,\Obhtwo,\,\OL)$ for \olcdm.

Validated against DESI DR2 BAO combined with the corresponding SPA chains, the compressed likelihoods reproduce the marginalized constraints and two-dimensional contour shapes from the full CMB analysis with good-to-excellent accuracy across all five model classes. The agreement is essentially exact in \olcdm, where all listed mean shifts relative to full-CMB analysis results are below about $0.13\sigma$, and very good in \wowacdm, \nulcdm, and \nuwowacdm. In \wowacdm, all listed mean shifts are below about $0.3\sigma$; in \nulcdm, the shifts in $\Om$ and $H_0$ are below $0.12\sigma$ while the neutrino-mass upper limit is about $13\%$ weaker; and in \nuwowacdm, the dark-energy and geometric parameters agree within $0.09\sigma$, with the $\sumnu$ mean shifted by about $0.4\sigma$ and its 95\% upper interval about $14\%$ broader. 

In \lcdm, we observe a small residual shift along the $\Om$--$H_0$ degeneracy (about $0.14\kmsMpc$ in $H_0$ and $\sim 2\times10^{-3}$ in $\Om$); this bias is likely due to the imperfect modeling of the CMB lensing signal with a 3-parameter compression and dark-energy parameterization that does not allow decoupling of the early- and late-universe physics. In \nuwowacdm, the neutrino-mass constraint is somewhat broader in the compressed run than in the full-CMB reference, consistent with the reduced leverage that \cmbfournuw\ carries on $\sumnu$ when dark-energy degrees of freedom are simultaneously freed, but the qualitative structure of the posterior is faithfully reproduced.  Overall, in all of the model classes that we studied, the compressed results on the late-time parameters describing dark energy are essentially unbiased relative to those from full-CMB chains.

As a practical guideline, users should employ the compressed likelihood corresponding to the physical extension under consideration. The \cmbthreew\, likelihood is appropriate for models whose effects are confined to the late-time expansion history and can be adequately described through modifications of the background evolution, while preserving the standard early-Universe physics, recombination history, and primary-CMB perturbation sector. The \cmbfourk\, and \cmbfournu\, likelihoods should be used for analyses in which spatial curvature or the summed neutrino mass are varied. Extensions that introduce new early-time physics, modify recombination, alter the primordial perturbation sector, significantly change the growth or lensing observables beyond the calibration domain, or otherwise lie outside the validated parameter space should instead be analyzed using the full CMB likelihood or a dedicated recalibration of the compressed likelihood. The mean vectors and covariance matrices required for implementation are collected in Table~\ref{tab:dvs_and_cov}.

% Our compressed likelihoods are particularly well-suited for rapid phenomenological studies of models that modify the late-time expansion history while leaving the early-Universe physics, recombination history, and primary-CMB perturbation sector unchanged. This is the regime in which the dominant CMB information entering low-redshift geometric inference is captured by the distance to last scattering, the acoustic scale, and the relevant physical-density variables. Models that directly alter recombination, the primordial perturbation sector, the radiation content outside the calibrated parameter space, or the growth/lensing sector would require a separate recalibration and validation of the compressed data vector and covariance. The mean vectors and covariance matrices required for implementation are collected in Table~\ref{tab:dvs_and_cov}.

\vspace{1cm}

\section{Code Availability}
\label{sec:code}

All data vectors, covariance matrices, likelihoods, and inference code used
in this paper are publicly available in the \textsc{CMBComp} repository at
\url{https://github.com/amoghsriv/CMBcomp}. Although the analyses presented
here were performed using CAMB and Cobaya, reproducing any of the results in
this paper does not require either package. The repository was designed to
provide a lightweight, standalone implementation of the compressed-CMB
likelihoods so that they can be used by researchers, students, or educators
without adopting a full cosmological parameter-inference framework.

The repository is self-contained and requires no Boltzmann solver or
inference framework. It provides a from-scratch implementation of the
cosmological background evolution, including photon and massive-neutrino
energy densities, CPL dark energy, spatial curvature, and the comoving sound
horizon calculation. The implementation is validated against CAMB to better
than $0.1\%$ in both $r_s$ and $\ell_a$ across all five model classes
considered in this work.

The five compressed-CMB likelihoods (\cmbthree, \cmbthreew, \cmbfournu,
\cmbfournuw, \cmbfourk) are implemented as standalone Gaussian likelihood
modules. The repository also includes a DESI DR2 BAO likelihood based on the
13-point $D_M/r_d$, $D_H/r_d$, and $D_V/r_d$ measurements, including the full
inter-tracer covariance and the \citet{Aubourg:2014yra} fitting formula for
the sound horizon at the drag epoch.

Parameter inference is performed using a custom adaptive
Metropolis--Hastings sampler with reflecting prior boundaries, an evolving
proposal covariance matrix, and online Gelman--Rubin convergence monitoring.
As a result, the complete analysis pipeline can be run in a standard Python
environment using only \textsc{NumPy}, \textsc{SciPy}, and \textsc{GetDist}.

Five annotated Jupyter notebooks, one for each cosmological model considered
in this paper, reproduce all validation figures and posterior constraints
presented here. These notebooks provide end-to-end examples that generate
chains directly from the compressed likelihoods without requiring CAMB,
Cobaya, CosmoMC, MontePython, or any other external cosmological inference
framework. Chains are written in GetDist format and can therefore be compared
directly with external full-CMB analyses.

A suite of unit tests validates the background calculations against known
reference values, and a parameter-ranges file is automatically written
alongside each chain so that GetDist's boundary-correction kernel density
estimation is applied consistently at hard prior limits (e.g.\ $\sum m_\nu
\geq 0$ and $w_a \geq -3$).

A user wishing to apply any of the compressions to a new cosmological model
or external dataset need only hard-code the relevant row of
Table~\ref{tab:dvs_and_cov} and instantiate the provided
\texttt{CompressedCMBLikelihood} class. No Boltzmann solver, likelihood
framework, or MCMC package is required. The accompanying
\texttt{DESIBAOLikelihood} class, background-cosmology module, and adaptive
sampler together provide a complete standalone inference pipeline, while
remaining fully compatible with CAMB/Cobaya-based workflows for users who
prefer them.

\begin{acknowledgments}
PB and DH acknowledge support from Department of Energy under contract DE-SC009193, the Leinweber Center for Theoretical Physics, and University of Michigan. We thank Cristhian Garcia-Quintero for helpful comments and suggestions on the manuscript.
\end{acknowledgments}

\bibliographystyle{apsrev4-2}
\bibliography{references}

@article{DESI:2025zgx,
    author = "Abdul-Karim, M. and others",
    collaboration = "DESI",
    title = "{{DESI} DR2 Results II: Measurements of Baryon Acoustic Oscillations and Cosmological Constraints}",
    eprint = "2503.14738",
    archivePrefix = "arXiv",
    primaryClass = "astro-ph.CO",
    journal = "Phys. Rev. D",
    volume = "112",
    number = "8",
    pages = "083515",
    year = "2025"
}

@article{Bansal:2025xxx,
    author = "Bansal, Prakhar and Huterer, Dragan",
    title = "{Expansion-history preferences of {DESI} {DR2} and external data}",
    eprint = "2502.07185",
    archivePrefix = "arXiv",
    primaryClass = "astro-ph.CO",
    journal = "Phys. Rev. D",
    volume = "112",
    number = "2",
    pages = "023528",
    year = "2025"
}

@article{Planck_PR3,
    author = "Aghanim, N. and others",
    collaboration = "Planck",
    title = "{{Planck} 2018 results. V. CMB power spectra and likelihoods}",
    eprint = "1907.12875",
    archivePrefix = "arXiv",
    primaryClass = "astro-ph.CO",
    doi = "10.1051/0004-6361/201936386",
    journal = "Astron. Astrophys.",
    volume = "641",
    pages = "A5",
    year = "2020"
}

@article{ACT:2023kun,
    author = "Madhavacheril, Mathew S. and others",
    collaboration = "ACT",
    title = "{The Atacama Cosmology Telescope: {DR6} Gravitational Lensing Map and Cosmological Parameters}",
    eprint = "2304.05203",
    archivePrefix = "arXiv",
    primaryClass = "astro-ph.CO",
    reportNumber = "FERMILAB-PUB-23-206-PPD",
    doi = "10.3847/1538-4357/acff5f",
    journal = "Astrophys. J.",
    volume = "962",
    number = "2",
    pages = "113",
    year = "2024"
}

@article{SPT-3G:2025bzu,
    author = "Camphuis, E. and others",
    collaboration = "SPT-3G",
    title = "{SPT-3G D1: CMB temperature and polarization power spectra and cosmology from 2019 and 2020 observations of the SPT-3G main field}",
    eprint = "2506.20707",
    archivePrefix = "arXiv",
    primaryClass = "astro-ph.CO",
    reportNumber = "FERMILAB-PUB-25-0144-PPD",
    doi = "10.1103/7wt3-9v2y",
    journal = "Phys. Rev. D",
    volume = "113",
    number = "8",
    pages = "083504",
    year = "2026"
}

@article{Bond:1997wr,
    author = "Bond, J. R. and Efstathiou, G. and Tegmark, M.",
    title = "{Forecasting cosmic parameter errors from microwave background anisotropy experiments}",
    eprint = "astro-ph/9702100",
    archivePrefix = "arXiv",
    reportNumber = "IASSNS-AST-97-10",
    doi = "10.1093/mnras/291.1.L33",
    journal = "Mon. Not. Roy. Astron. Soc.",
    volume = "291",
    pages = "L33--L41",
    year = "1997"
}

@article{Wang:2007mza,
    author = "Wang, Yun and Mukherjee, Pia",
    title = "{Observational Constraints on Dark Energy and Cosmic Curvature}",
    eprint = "astro-ph/0703780",
    archivePrefix = "arXiv",
    doi = "10.1103/PhysRevD.76.103533",
    journal = "Phys. Rev. D",
    volume = "76",
    pages = "103533",
    year = "2007"
}

@article{Mukherjee:2008kd,
    author = "Mukherjee, Pia and Kunz, Martin and Parkinson, David and Wang, Yun",
    title = "{Planck priors for dark energy surveys}",
    eprint = "0803.1616",
    archivePrefix = "arXiv",
    primaryClass = "astro-ph",
    doi = "10.1103/PhysRevD.78.083529",
    journal = "Phys. Rev. D",
    volume = "78",
    pages = "083529",
    year = "2008"
}

@article{Elgaroy:2007bv,
    author = "Elgaroy, Oystein and Multamaki, Tuomas",
    title = "{On using the CMB shift parameter in tests of models of dark energy}",
    eprint = "astro-ph/0702343",
    archivePrefix = "arXiv",
    doi = "10.1051/0004-6361:20077292",
    journal = "Astron. Astrophys.",
    volume = "471",
    pages = "65",
    year = "2007"
}

@article{Vonlanthen:2010cd,
    author = "Vonlanthen, Marc and Rasanen, Syksy and Durrer, Ruth",
    title = "{Model-independent cosmological constraints from the CMB}",
    eprint = "1003.0810",
    archivePrefix = "arXiv",
    primaryClass = "astro-ph.CO",
    doi = "10.1088/1475-7516/2010/08/023",
    journal = "JCAP",
    volume = "2010",
    number = "08",
    month = "08",
    pages = "023",
    year = "2010"
}

@article{Huang:2015vpa,
    author = "Huang, Qing-Guo and Wang, Ke and Wang, Shuang",
    title = "{Distance priors from Planck 2015 data}",
    eprint = "1509.00969",
    archivePrefix = "arXiv",
    primaryClass = "astro-ph.CO",
    doi = "10.1088/1475-7516/2015/12/022",
    journal = "JCAP",
    volume = "2015",
    number = "12",
    pages = "022",
    year = "2015"
}

@article{Zhai:2019nad,
    author = "Zhai, Zhongxu and Park, Chan-Gyung and Wang, Yun and Ratra, Bharat",
    title = "{CMB distance priors revisited: effects of dark energy dynamics, spatial curvature, primordial power spectrum, and neutrino parameters}",
    eprint = "1912.04921",
    archivePrefix = "arXiv",
    primaryClass = "astro-ph.CO",
    doi = "10.1088/1475-7516/2020/07/009",
    journal = "JCAP",
    volume = "2020",
    number = "07",
    pages = "009",
    year = "2020"
}

@article{Lewis:1999bs,
    author = "Lewis, Antony and Challinor, Anthony and Lasenby, Anthony",
    title = "{Efficient computation of CMB anisotropies in closed FRW models}",
    journal = "Astrophys. J.",
    volume = "538",
    year = "2000",
    pages = "473-476",
    doi = "10.1086/309179",
    eprint = "astro-ph/9911177",
    archivePrefix = "arXiv",
    primaryClass = "astro-ph"
}

@article{Torrado:2020dgo,
    author = "Torrado, Jes\'us and Lewis, Antony",
    title = "{Cobaya: code for Bayesian analysis of complex models}",
    eprint = "2005.05290",
    archivePrefix = "arXiv",
    primaryClass = "astro-ph.IM",
    doi = "10.1088/1475-7516/2021/05/057",
    journal = "JCAP",
    volume = "2021",
    number = "05",
    pages = "057",
    year = "2021"
}

@article{Lewis:2019xzd,
    author = "Lewis, Antony",
    title = "{{GetDist}: a Python package for analysing Monte Carlo samples}",
    eprint = "1910.13970",
    archivePrefix = "arXiv",
    primaryClass = "astro-ph.IM",
    doi = "10.1088/1475-7516/2025/08/025",
    journal = "JCAP",
    volume = "2025",
    number = "08",
    pages = "025",
    year = "2025"
}

@article{Chevallier:2000qy,
    author = "Chevallier, Michel and Polarski, David",
    title = "{Accelerating universes with scaling dark matter}",
    eprint = "gr-qc/0009008",
    archivePrefix = "arXiv",
    doi = "10.1142/S0218271801000822",
    journal = "Int. J. Mod. Phys. D",
    volume = "10",
    pages = "213--224",
    year = "2001"
}

@article{Linder:2002et,
    author = "Linder, Eric V.",
    title = "{Exploring the expansion history of the universe}",
    eprint = "astro-ph/0208512",
    archivePrefix = "arXiv",
    doi = "10.1103/PhysRevLett.90.091301",
    journal = "Phys. Rev. Lett.",
    volume = "90",
    pages = "091301",
    year = "2003"
}

@article{SPT-3G:2022hvq,
    author = "Balkenhol, L. and others",
    collaboration = "SPT-3G",
    title = "{Measurement of the CMB temperature power spectrum and constraints on cosmology from the SPT-3G 2018 TT, TE, and EE dataset}",
    eprint = "2212.05642",
    archivePrefix = "arXiv",
    primaryClass = "astro-ph.CO",
    doi = "10.1103/PhysRevD.108.023510",
    journal = "Phys. Rev. D",
    volume = "108",
    number = "2",
    pages = "023510",
    year = "2023"
}

@article{Aubourg:2014yra,
    author       = {Aubourg, \'Eric and others},
    title        = {Cosmological implications of baryon acoustic oscillation
                    measurements},
    journal      = {Phys.\ Rev.\ D},
    volume       = {92},
    pages        = {123516},
    year         = {2015},
    doi          = {10.1103/PhysRevD.92.123516},
    eprint       = {1411.1074},
    archivePrefix= {arXiv},
    primaryClass = {astro-ph.CO},
}

\end{document}